\documentclass[aps,prl,showpacs,twocolumn,superscriptaddress]{revtex4}
\usepackage{amsmath}
\usepackage{amssymb}
\usepackage{epsfig}
\usepackage{color}
\usepackage{graphicx,amsmath}

\begin{document}
\title{Comment on Zeeman-Driven Lifshitz Transition: A Model for the Experimentally Observed
Fermi-Surface Reconstruction in $\rm YbRh_2Si_2$ (A. Hackl and M.
Vojta, \prl {\bf 106}, 137002 (2011))}
\author{V. R. Shaginyan} \affiliation{Petersburg
Nuclear Physics Institute,  Gatchina, 188300,
Russia}\affiliation{Clark Atlanta University, Atlanta, GA 30314,
USA} \author{M. Ya. Amusia}\affiliation{Racah Institute of Physics,
Hebrew University, Jerusalem 91904, Israel}\affiliation{Ioffe
Physical Technical Institute, RAS, St. Petersburg 194021, Russia}
\author{J.~W.~Clark}
\affiliation{McDonnell Center for the Space Sciences \& Department
of Physics, Washington University, St.~Louis, MO 63130, USA}
\author{A. Z. Msezane}\affiliation{Clark Atlanta University, Atlanta, GA 30314,
USA}\author{K. G. Popov}\affiliation{Komi Science Center, Ural
Division, RAS, Syktyvkar, 167982, Russia} \author{M. V. Zverev}
\affiliation{Russian Research Center Kurchatov Institute, Moscow,
123182, Russia}\affiliation{ Moscow Institute of Physics and
Technology, Moscow, 123098, Russia}
\author{V. A. Khodel}
\affiliation{Russian Research Center Kurchatov Institute, Moscow,
123182, Russia} \affiliation{McDonnell Center for the Space
Sciences \& Department of Physics, Washington University,
St.~Louis, MO 63130, USA}


\maketitle

The letter \cite{vojta} (L) presents a model description of the
non-Fermi-liquid (NFL) behavior of the heavy-fermion (HF) metal
$\rm YbRh_2Si_2$ (YRS) at a quantum critical point (QCP) that can
be tuned by imposition of an extremely weak external magnetic field
$B$. Conventional wisdom assumes that behavior in the vicinity of
the QCP is described within a Kondo-breakdown scenario driven by
quantum criticality \cite{steglich}. However, applications of this
scenario fail to account for experimental facts. The proposed model
\cite{vojta}, based on the idea of a Lifshitz topological
transition \cite{volrev} occurring in a strongly correlated
electron liquid, can explain several striking qualitative features
of NFL behavior of YRS, such as maxima at temperatures $T_{\rm
max}$ in both the susceptibility $\chi$ and specific heat
coefficient $\gamma$ at the crossover temperature $T^*$ with
$T_{\rm max}\propto B$.

In this Comment we confirm that the explanation of these striking
features of NFL behavior of YRS given in L is qualitatively correct
and in accord with studies \cite{volrev,ckz,pr,khodb,schofield},
but we argue that the main quantitative results of L are incorrect
and may misdirect the reader. In L the authors base their study on
a toy model of free fermions with a sample DOS independent of
temperature $T$, for which application of the magnetic field
induces the Zeeman-driven Lifshitz transition and leads to the
aforementioned features. Since the Zeeman splitting generates a
universal behavior of YRS, the qualitative results of L are valid.
This universal behavior includes the existence of such regimes as
the Landau Fermi liquid, the crossover, and the NFL behavior, with
$\chi$ and $\gamma$ exhibiting maxima in the crossover region. The
known relevance of the Lifshitz topological scenario to the NFL
effects in YRS associated with imposition of a magnetic field $B$
date back to \cite{schofield,ckz,pr,khodb}. On the other hand, the
main quantitative results of L are incorrect because the authors
ignore a key relationship between the DOS and the Fermi function
that ensures $T$-dependence of the DOS. These quantities are linked
to each other by the Landau equation in terms of the interaction
function. If this link is disconnected, the analysis becomes
inconsistent \cite{khodb,ckz,pr}.  We note that Fig.~55 \cite{pr}
shows the strong dependence of the DOS on $T$. This same dependence
was observed in Ref. \cite{shim} and recently in Ref. \cite{ernst},
Fig.~3c. Consequently, the quantitative results of L related to
$\chi$, $\gamma$, and the Hall coefficient and the DOS structure at
low temperatures are unreliable. Thus, the readers are likely to be
confused by some conclusions of L, especially the statement
``Further experiments to discriminate the different scenarios are
proposed,'' for the conclusions are not intrinsic to the proposed
in L quasiparticle scenario but rather artifacts introduced by the
toy model. For example, a smeared jump in the Hall coefficient
dictated by the toy model is at variance with the recent
demonstration that the Hall coefficient undergoes a sharp jump
discontinuity \cite{steg}, while the same scenario is in agreement
with the facts \cite{pr}.

\end{document}